\documentclass[10pt,twocolumn,twoside]{IEEEtran}

\usepackage[numbers,sort&compress]{natbib}
\usepackage[pdftex]{graphicx}
\graphicspath{{./}{graphics/},{code/}}
\usepackage{subcaption}
\usepackage{NotationMacros}
\usepackage{ScatteringFDNMacros}
\usepackage{lipsum}
\usepackage{chngpage}
\usepackage{booktabs}
\newcommand{\ra}[1]{\renewcommand{\arraystretch}{#1}}
\usepackage[hidelinks]{hyperref}

\renewcommand{\schlecsn}[1]{#1}

\begin{document}

\title{Scattering in Feedback Delay Networks}

\author{S.J. Schlecht, \IEEEmembership{Senior Member, IEEE}, and E.A.P. Habets, \IEEEmembership{Senior Member, IEEE} 
\thanks{S.J. Schlecht is with the Acoustics Laboratory, Department of Signal Processing and Acoustics, Aalto University, FI-00076 Espoo, Finland (e-mail: sebastian.schlecht@aalto.fi); E.A.P. Habets is with the International Audio Laboratories Erlangen (a joint institution of the University of Erlangen-Nuremberg and Fraunhofer IIS), Germany (e-mail: emanuel.habets@audiolabs-erlangen.de).}
}

\markboth{IEEE TRANSACTIONS ON Audio, Speech, and Language Processing,~Vol.~0, No.~0, June~2020}%
{Schlecht and Habets: Scattering in Feedback Delay Networks}

\maketitle

\IEEEpeerreviewmaketitle
\begin{abstract}
Feedback delay networks (FDNs) are recursive filters, which are widely used for artificial reverberation and decorrelation. One central challenge in the design of FDNs is the generation of sufficient echo density in the impulse response without compromising the computational efficiency. In a previous contribution, we have demonstrated that the echo density of an FDN can be increased by introducing so-called delay feedback matrices where each matrix entry is a scalar gain and a delay. In this contribution, we generalize the feedback matrix to arbitrary lossless filter feedback matrices (FFMs). As a special case, we propose the velvet feedback matrix, which can create dense impulse responses at a minimal computational cost. Further, FFMs can be used to emulate the scattering effects of non-specular reflections. We demonstrate the effectiveness of FFMs in terms of echo density and modal distribution.  
\end{abstract}


\section{Introduction}
\label{sec:Intro}

\begin{figure}[!tb]
\centering
\includegraphics[width=0.49\textwidth]{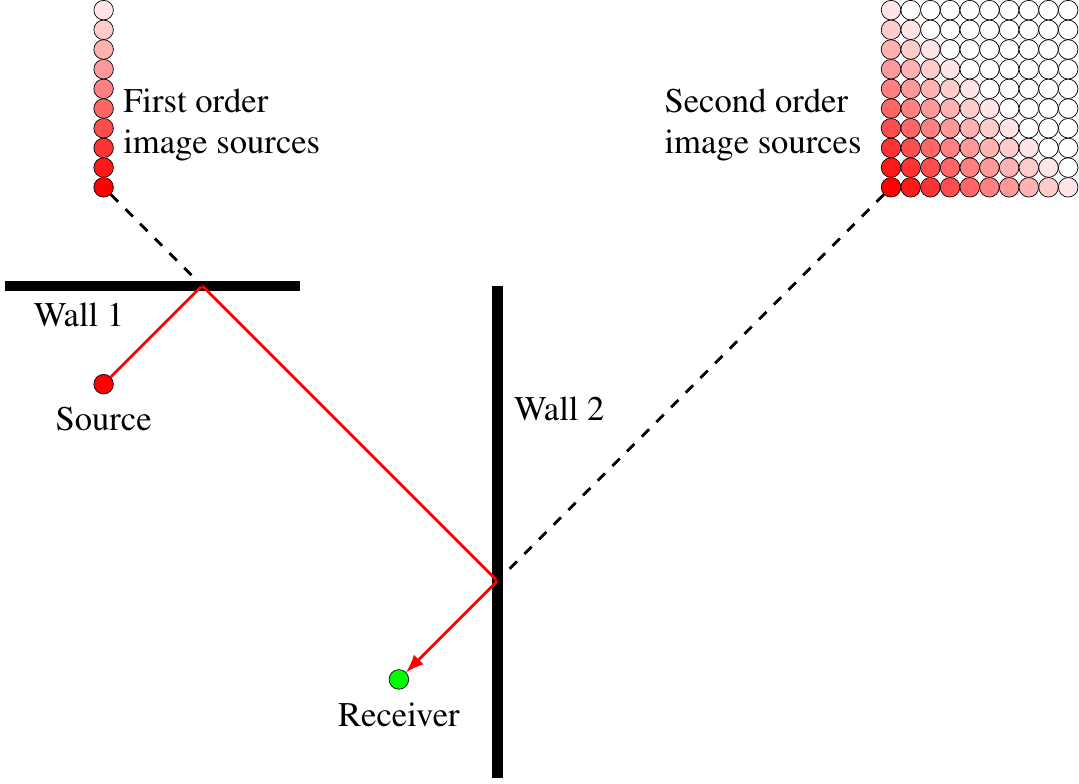}
\caption{Second-order reflections at two rough surfaces are represented with an equivalent set of image sources (inspired by \cite{Siltanen:2012ja}). The color saturation indicates the energy of the image sources.}
\label{fig:scattering}
\end{figure}

\IEEEPARstart{I}{f} a sound is emitted in a room, the sound waves travel through space and are repeatedly reflected at the room boundaries resulting in acoustic reverberation \cite{Kuttruff:2009vl}. If the sound is reflected at a smooth boundary, the reflection is coherent (specular), while it is incoherent (scattered) when reflected by a rough surface. In small to large rooms, the first-order specular reflections arrive between 10 - 100~ms, whereas the time scale of incoherent reflections is a few milliseconds \cite{Siltanen:2012ja}. In geometric room acoustics, an incoherent reflection can be effectively generated from a set of closely spaced image sources \cite{Siltanen:2012ja}, see Fig.~\ref{fig:scattering}. Consequently, each of the primary image sources (specular) is spread in time, and therefore the resulting echo density increases while the number of image sources still follows a polynomial of degree three \cite{Kuttruff:2009vl}. In this work, we propose a method to introduce scattering-like effects into artificial reverberation filter structures effectively.

Many artificial reverberators have been developed in recent years \cite{Valimaki:2012jv}, among which the feedback delay network (FDN), initially proposed by Gerzon \cite{Gerzon:1971tu} and further developed in \cite{Jot:1991tq, Rocchesso:1997ct}, is one of the most popular. The FDN consists of $\systemOrder$ delay lines combined with attenuation filters, which are fed back via a scalar feedback matrix $\Fbm$. 
A significant challenge of FDN design is to achieve sufficient echo density because of an inherent trade-off between three aspects: computational complexity, mode density, and echo density. A higher number of delays increases both modal and echo density, but also the computational complexity. 
To reproduce a scattering-like effect, we require a set of long delays that are proportional to the mean free path \cite{Rocchesso:1995is, Schlecht:2017il}, and a set of filters to add the short-term density to each reflection (see Fig.~\ref{fig:scattering}). Feedforward-feedback allpass filters have been introduced with the delay lines to increase the short-term echo density \cite{Schroeder:1961ke, Moorer:1979hi}. Alternatively, allpass filters may be placed after the delay lines \cite{Vaananen:1997wj, Dahl:2000tz}, which in turn doubles the effective size of the FDN \cite{Schlecht:2015hi}.
Alternatively, scattering filters can be introduced in series to the FDN. The post-filtering is, however, not optimal because of the perceivable repetitions and an elaborate time-varying switching was proposed to overcome this problem \cite{Lee:2012uw}.     

Towards a possible solution, we introduced in \cite{Schlecht:2019tl} the delay feedback matrix (DFM), where each matrix entry is a scalar gain and a delay. In this contribution, we generalize the feedback matrix of the FDN to a filter feedback matrix (FFM), which then results in a filter feedback delay network (FFDN). As a special case of the FFM, we present the velvet feedback matrix (VFM), which can create ultra-dense impulse responses at a minimal computational cost. The VFM is inspired by the work of Karjalainen, V\"alim\"aki et al. \cite{Jarvelainen:2007wp, HolmRasmussen:2013uc, Lee:2012uw, Valimaki:2012jv} in which velvet noise sequences, i.e., sparse time-domain responses with few $\pm1$ pulses, have been proposed to create perceptually white noise. Due to the general formulation, the proposed FFM is directly applicable for related techniques such as digital waveguides \cite{SmithIII:1992bn}, waveguide web \cite{Stevens:2017kj}, scattering delay networks \cite{DeSena:2015bb} and directional FDNs \cite{Alary:2019tr}.
 
The remainder of this work is organized as follows. In Section~\ref{sec:priorArt}, we review the general FDN structure, conditions for losslessness as well as prior techniques for introducing scattering-like effects. In Section~\ref{sec:FFM}, we introduce a complete characterization of the proposed filter feedback matrices alongside various special cases, including the VFM. In Section~\ref{sec:modalDistribution}, we study the effect the FFM has on the modal distribution. In Section~\ref{sec:temporalDensity}, we demonstrate the echo density of the proposed FFDN.

\begin{figure}[!tb]
\centering
\includegraphics[width=0.5\textwidth]{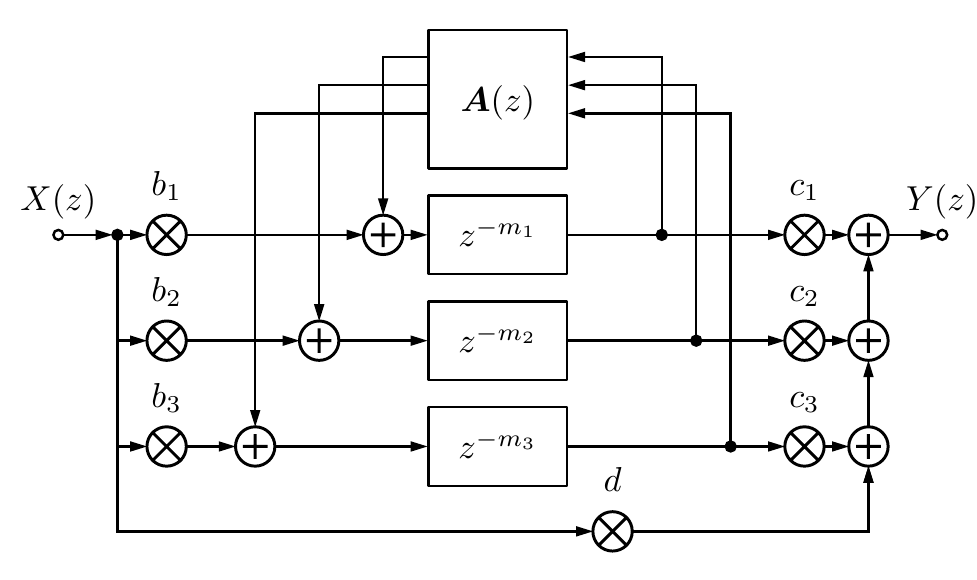}
\caption{Proposed filter feedback delay network (FFDN) with three delays, i.e., $\matSize = 3$, and a filter feedback matrix (FFM) $\Fbm(z)$ instead of the standard scalar feedback matrix $\Fbm$. }
\label{fig:paraunitaryFDN}
\end{figure}


\section{Filter Feedback Delay Network}
\label{sec:priorArt}

In the following, we present the proposed FFM extension resulting in the filter feedback delay network (FFDN) and the corresponding conditions for losslessness and homogeneous loss.

\subsection{ FFDN Structure }
The proposed FFDN is similarly structured as the standard FDN, but employs a filter feedback matrix (FFM) $\Fbm(z)$ instead of a scalar feedback matrix $\Fbm$ (see Fig.~\ref{fig:paraunitaryFDN}). The transfer function of the FFDN is
\begin{equation}
	H(z) = \Outgain\tran \bracket{\stdDelayArg{z^{-1}} - \Fbm(z)}^{-1} \Ingain + \directgain ,
	\label{eq:PFDNtrans}
\end{equation}
where the column vectors $\Ingain$, $\Outgain$, and, scalar $\directgain$ denote the input, output and direct gains, respectively. The lengths of the $\matSize$ delays in samples are given by $\Delay = [\delay_1, \dots, \delay_\matSize] \in \set{N}^\matSize$. $\stdDelay =\diag{[z^{-\delay_1}, z^{-\delay_2},\dots,z^{-\delay_\matSize} ]}$ is the diagonal $\matSize \times \matSize$ delay matrix.  

The FFM may consist of finite and infinite impulse response (FIR and IIR) filters. However, in this work, we largely focus on FIR FFMs. An FIR FFM $\Fbm(z)$ may be expressed in terms of the scalar coefficient matrices $\Fbm_0, \Fbm_1, \dots, \Fbm_\filterOrder$, i.e., 
\begin{equation}
	\Fbm(z) = \Fbm_0 + z^{-1} \Fbm_1 + z^{-2} \Fbm_2 + \dots + z^{-L} \Fbm_\filterOrder,
	\label{eq:FIRFFM}
\end{equation}
where $\filterOrder$ is the maximum filter order of $\Fbm(z)$ if $\Fbm_\filterOrder \neq \mat{0}$. \schlecsn{The length of the FIR filters $\filterOrder$ relates to the temporal spreading caused by the scattered reflection at rough boundaries.} Every matrix element $\fbm_{ij}(z)$ is an FIR filter with filter order at most $\filterOrder$. \schlecsn{ The McMillan degree of the FIR FFM $\deg \Fbm(z)$ is the minimum number of required delay elements to implement $\Fbm(z)$.}

Thus, the minimal number of delay elements to realize the proposed FFDN in \eqref{eq:PFDNtrans} is 
\begin{equation}
	\N = \deg \Fbm(z) + \sum_{\iterIndex = 1}^\matSize \delay_\iterIndex.
	\label{eq:totalSystemOrder}
\end{equation}
To achieve a scattering-like effect as discussed in the introduction, we choose the FFM order $\filterOrder$ relatively small compared to the main delays $\Delay$. In Sections~\ref{sec:modalDistribution} and \ref{sec:temporalDensity}, more details on practical designs are provided.

\begin{figure*}[tb]
\centering
\includegraphics[width=\textwidth]{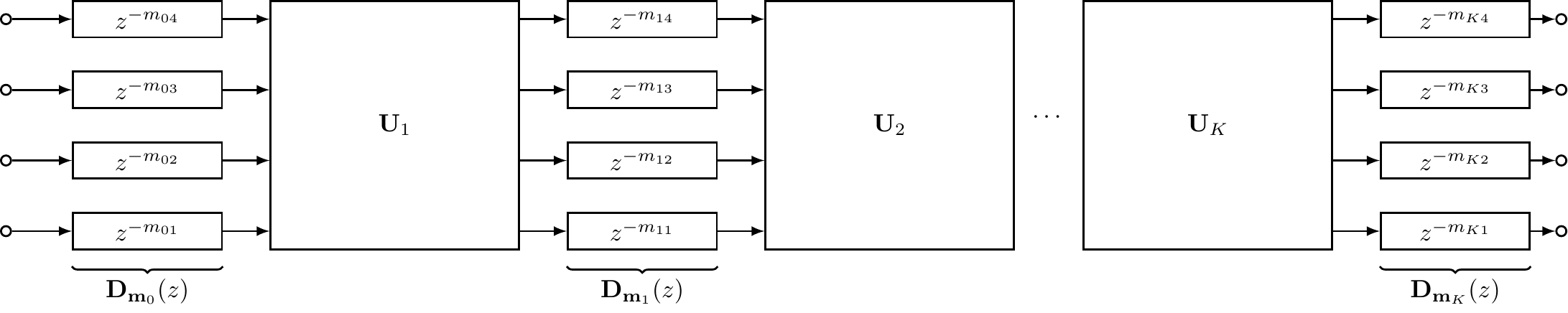}
\caption{General structure of a FIR filter feedback matrix (FFM) as defined in \eqref{eq:FFMfactorloose} with $\matSize = 4$ and $K$ stages.}
\label{fig:paraunitaryGeneralFIR}
\end{figure*}

\subsection{Losslessness}
FDNs are commonly designed as lossless systems, i.e., all system poles lie on the unit circle. The poles of the proposed FFDN are the eigenvalues of the polynomial matrix $\P = \stdDelayArg{z^{-1}} - \Fbm(z)$, which in turn are the roots of the generalized characteristic polynomial (GCP) \cite{Schlecht:2019uj}, i.e., 
\begin{equation}
\gcp(z) = z^{\deg \Fbm(z)} \detp{\P}.	
\end{equation}
The extra factor $z^{\deg \Fbm(z)}$ is necessary to turn $\detp{\P}$ into a proper polynomial. The lossless property of general unitary-networks, which in particular applies to the proposed FFDN, was described by Gerzon \cite{Gerzon:1976fm}. An FFDN is lossless if $\Fbm(z)$ is paraunitary, i.e., $\Fbm(\inv{z})\herm \Fbm(z) = \eye $, where $\eye$ is the identity matrix and $\cdot\herm$ denotes the complex conjugate transpose \cite{Gerzon:1976fm}. Although also non-paraunitary FFMs may yield lossless FFDNs \cite{Schlecht:2017jt,Schlecht:2019tl}, here we focus on paraunitary FFMs only. Paraunitary matrices are particularly useful as they are closed under multiplication, i.e., if $\mat{A}(z)$ and $\mat{B}(z)$ are paraunitary, then $\mat{A}(z)\mat{B}(z)$ is paraunitary as well \cite{Vaidyanathan1990}. 
\schlecsn{The McMillan degree of a causal lossless $\mat{A}(z)$ is given by the polynomial degree of the matrix determinant \cite[p.757]{Vaidyanathan:1993uh}
\begin{equation}
	\deg \Fbm(z) = \deg \paren{\detp{\Fbm(z)}}.
\end{equation} 
An efficient FFT-based method for determining the polynomial matrix determinant $\detp{\Fbm(z)}$ is given in \cite{Hromcik:1999uu}. }

\subsection{Lossy Feedback}  
\label{sec:lossyFeedback}
Homogeneous loss is introduced into a lossless FFDN by replacing each delay element $\inv{z}$ with a lossy delay filter $\gain(z) \inv{z}$, where $\gain(z)$ is ideally zero-phase with a positive frequency-response. The frequency-dependent gain-per-sample $\gain(\ejw)$ relates to the resulting reverberation time $\RT(\omega)$ by 
\begin{equation}
	\gain(\ejw) = \frac{-60}{\FS \RT(\omega) },
	\label{eq:attenuationRT60}
\end{equation}
where $\FS$ is the sampling frequency and $\omega$ is the angular frequency \cite{Jot:1991tq}. However, as substitution with lossy delays is impractical, the attenuation filters are lumped into a single filter per delay line. Thus, the lossy FFM is
 \begin{equation}
	\Fbm(z) = \Unitary(z) \GainMat(z),
	\label{eq:lumpedAttenuation}
\end{equation}
where $\Unitary(z)$ is paraunitary and $\GainMat(z)$ a diagonal attenuation matrix. The attenuation matrix $\GainMat(z)$ is chosen such that the frequency-dependent eigenvalue magnitudes are similar for
\begin{equation}
	\stdDelayArg{ \frac{\gain(z)}{z} } - \Unitary \p*{\frac{z}{\gain(z)}} \textrm{ and } \stdDelayArg{z^{-1}} - \Unitary(z) \GainMat(z).
	\label{eq:lumpedAttenuationCriteria}
\end{equation}
In standard FDNs with a scalar feedback matrix $\Unitary$, \eqref{eq:lumpedAttenuationCriteria} yields the well-known delay-proportional attenuation \cite{Jot:1991tq}
\begin{equation}
	\abs{\GainMat(\ejw)} = \diag{{\gain(\ejw)}^{\Delay}}.
	\label{eq:attenuationFDN}
\end{equation}
In Section~\ref{sec:modalDistribution}, we give a general solution for FFMs. In the following section, we present various FFM designs.

%
%
%
%


%

\section{Filter Feedback Matrix}
\label{sec:FFM}

The filter feedback matrix (FFM) $\mat{A}(z)$ can be either realized by IIR or FIR filters. In the following, we discuss IIR FFMs and give a complete characterization of FIR FFMs. Then, we present particularly useful designs of FIR FFMs. We conclude this section by discussing the implementation and computational complexity.

\subsection{IIR FFMs}
The IIR FFM, i.e., each matrix element $\fbm_{ij}(z)$ is an IIR filter, is the most general FFM. IIR filters can produce long tails at little computational cost and therefore appear advantageous for increasing the density in artificial reverberation applications. In fact, allpass FDNs \cite{Vaananen:1997wj,Dattorro:1997up,Dahl:2000tz} are a special case of IIR FFMs with
\begin{equation}
	\Fbm(z) = \Unitary \, \Allpass(z),
\end{equation}
where $\Allpass(z)$ is a diagonal matrix of IIR allpass filters and $\Unitary$ is a scalar unitary matrix.
\schlecsn{Nonetheless, we do not consider further designs in this class and turn our attention to FIR FFMs, which can be parametrized and implemented highly efficiently.}

\begin{figure*}[!tb]
	\begin{subfigure}[b]{0.49\textwidth}
		\includegraphics[]{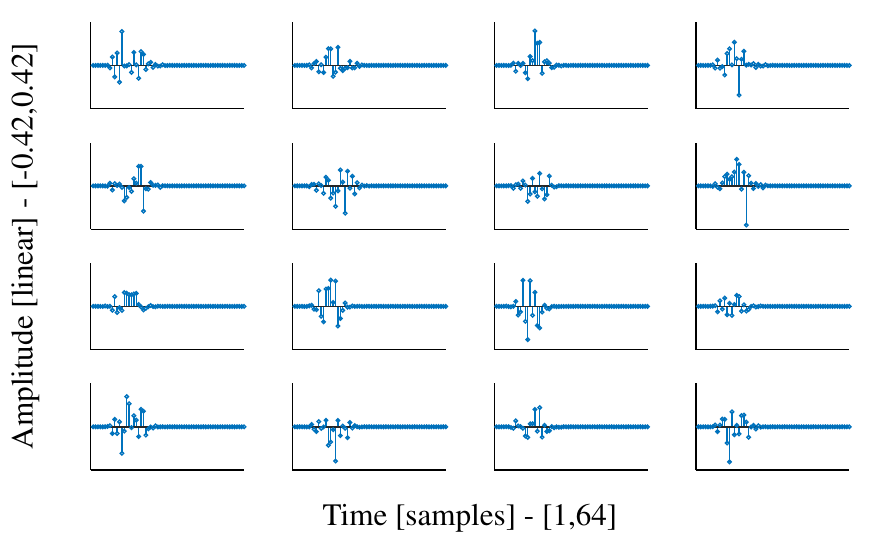}
		\caption{Elemental block feedback matrix (EBFM) with $\numStages = 64$ in \eqref{eq:FFMfactorization}.}
		\label{fig:FFM_elemental64}
	\end{subfigure}
	\hspace{0.3cm}
	\begin{subfigure}[b]{0.49\textwidth}
		\includegraphics[]{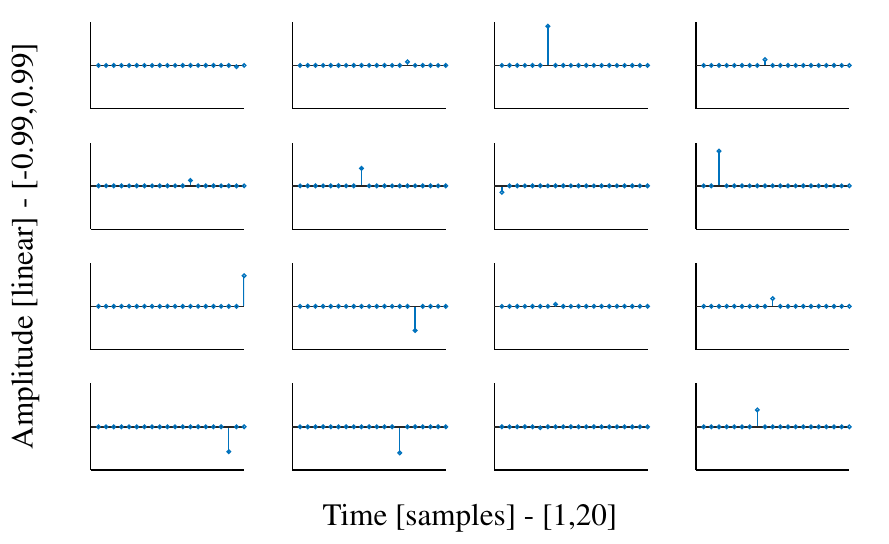}
		\caption{Delay feedback matrix (DFM) with $\numStages = 1$ in \eqref{eq:DFMdefine}.}
		\label{fig:FFM_delayMatrix}
	\end{subfigure}
	
	\begin{subfigure}[b]{0.49\textwidth}
		\includegraphics[]{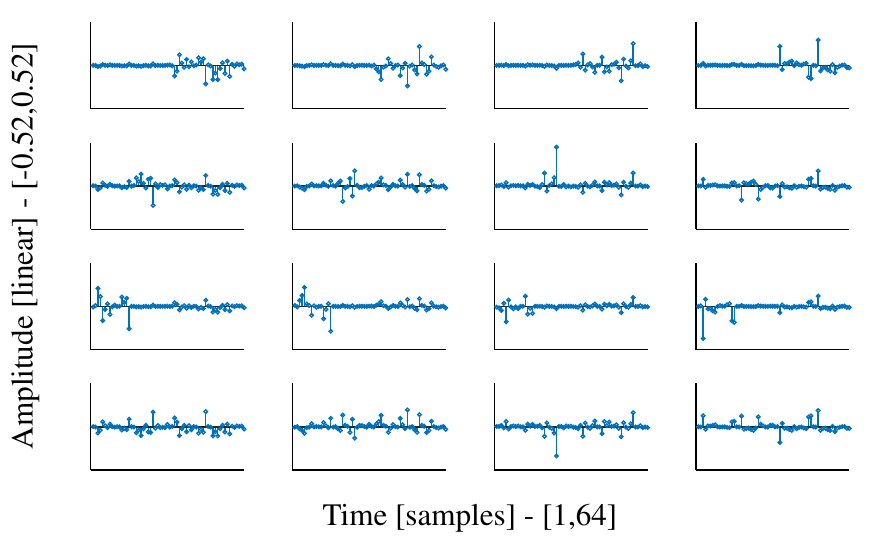}
		\caption{Random dense feedback matrix (RDFM) with $\numStages = 3$ in \eqref{eq:denseIteration}.}
		\label{fig:FFM_randomDense}
	\end{subfigure}
	\hspace{0.3cm}
	\begin{subfigure}[b]{0.49\textwidth}
		\includegraphics[]{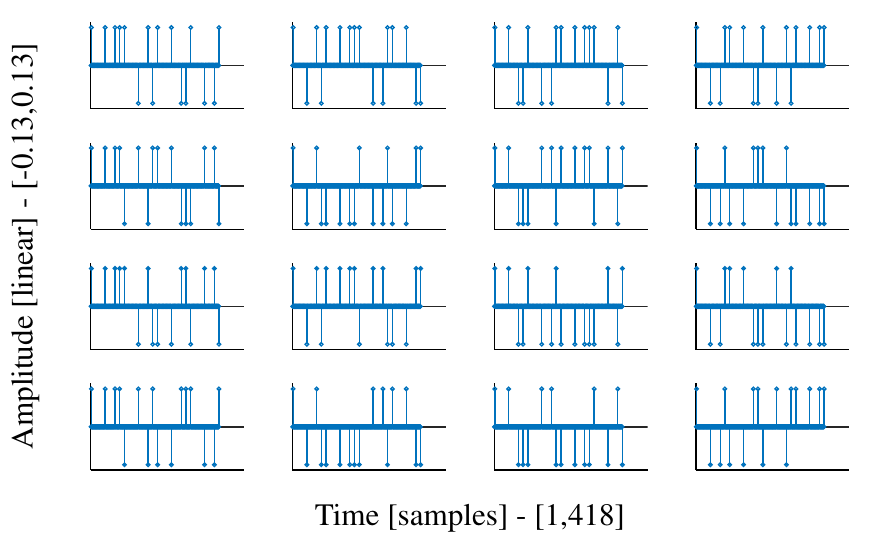}
		\caption{Velvet feedback matrix (VFM) with $\numStages = 2$ and $\velvetDensity = \frac{1}{30}$ in \eqref{eq:HadamardIteration}.}
		\label{fig:FFM_velvet}
	\end{subfigure}
	\caption{Paraunitary filter feedback matrices $\Fbm(z)$ with $\matSize = 4$. The subplots depict the filter coefficients of the matrix entries $\fbm_{ij}(z)$ with $1 \leq i,j \leq \matSize$. The pre- and post-delays $\Delay_0$ and $\Delay_\numStages$ are zero except for the delay feedback matrix in Fig.~\ref{fig:FFM_delayMatrix}.}
	\label{fig:FFM}
\end{figure*}

\subsection{Characterizing FIR FFMs}

Vaidyanathan gave a full characterization of paraunitary FIR FFMs in \cite{Vaidyanathan:1993uh}, which we reproduce in the following for completeness. The elemental building block is 
\begin{equation}
	\ElementFIR(z) = \eye - \elementFIR\elementFIR\herm + \inv{z} \elementFIR\elementFIR\herm, 
	\label{eq:elementalBlocks}
\end{equation}
where $\elementFIR$ is a $\matSize \times 1$ vector with unit norm, i.e., $\elementFIR\herm \elementFIR = 1$. \schlecsn{Note that in \eqref{eq:elementalBlocks}, the projection onto $\elementFIR$ is first subtracted and then added one sample later.} The degree of the elemental building block is one, i.e., $\deg \ElementFIR(z) = 1$. Due to the closure under multiplication, the matrix product
\begin{equation}
	\Fbm(z) = \ElementFIR_\numStages(z) \cdots \ElementFIR_2(z)  \ElementFIR_1(z) 
	\label{eq:FFMfactorization}
\end{equation}
with corresponding vectors $\elementFIR_1, \dots, \elementFIR_\numStages$ yields a paraunitary matrix. It can be shown \cite{Vaidyanathan:1993uh} that any FIR paraunitary matrix $\Fbm(z)$ of degree $\numStages$ can be factorized into the form \eqref{eq:FFMfactorization}.

A less rigorous, but more practical factorization of paraunitary FIR matrices is given by 
\begin{equation}
	\Fbm(z) = \stdDelayArgm{\Delay_\numStages}{z} \Unitary_\numStages \cdots  \Unitary_2  \stdDelayArgm{\Delay_1}{z} \Unitary_1 \stdDelayArgm{\Delay_0}{z},
	\label{eq:FFMfactorloose}
\end{equation}
where $\Unitary_1, \dots, \Unitary_\numStages$ are $\matSize \times \matSize$ unitary matrices and $\Delay_0, \Delay_1, \dots, \Delay_\numStages$ are vectors of $\matSize$ integer delays \cite[pp.~767]{Vaidyanathan:1993uh}\footnote{\schlecsn{In \cite[pp.~767]{Vaidyanathan:1993uh}, it is sufficient to choose the delays $\Delay_i = [0, \dots, 0, 1]$ and set $\Unitary_i$ as sequence of Givens rotations.}}. 
In this formulation, the FFM mainly introduces $\numStages$ delay and mixing stages within the main FDN loop (see Figure~\ref{fig:paraunitaryGeneralFIR}). A few examples of FIR FFMs are depicted in Fig.~\ref{fig:FFM}, and we explain their construction in the following.

\subsection{Elemental Block FIR FFMs}
The characterization in \eqref{eq:elementalBlocks} and  \eqref{eq:FFMfactorization} can be directly employed to create a paraunitary elemental block feedback matrix (EBFM) by choosing a set of vectors $\elementFIR_1, \dots, \elementFIR_\numStages$. Unfortunately, merely randomizing the vectors $\elementFIR_i$ does not yield satisfactory results as the energy of the FIR tends to concentrate in time. Figure~\ref{fig:FFM_elemental64} shows an example matrix with $\numStages = 64$, where all entries of the vectors $\elementFIR_i$ were picked from a Gaussian distribution and subsequently normalized. The concentration of energy is in direct opposition to the purpose of the FFM, namely to spread the energy of the reflection in time. We have tested various other strategies for choosing vectors $\elementFIR_i$ to no avail. In the following, we present alternative characterizations based on \eqref{eq:FFMfactorloose}, which results in more intuitive designs.

\subsection{Delay Feedback Matrix}
The delay feedback matrix (DFM) is a paraunitary FIR matrix, which was introduced by the present authors in \cite{Schlecht:2019tl}. It can be conveniently expressed in terms of \eqref{eq:FFMfactorloose} with a single stage, i.e, $\numStages = 1$:
\begin{equation}
	\Fbm(z) = \stdDelayArgm{\Delay_1}{z} \Unitary_1 \stdDelayArgm{\Delay_0}{z}.
	\label{eq:DFMdefine} 
\end{equation}
The implementation of the DFM matrix is similar to the unitary matrix $\Fbm$ except that the signal vector is read and written at dedicated points in the main delays, while all other arithmetic operations are unaltered. An example of the DFM is shown in Fig.~\ref{fig:FFM_delayMatrix}.

\subsection{Dense FFMs and Paraunitary Hadamard}
Through an iterative process, we compose \emph{dense} paraunitary FFMs, i.e., no matrix entry of any coefficient matrix $\Fbm_i$ in \eqref{eq:FIRFFM} is zero. For this, we introduce the notion of a \emph{(strictly) column-distinct} FFM, i.e.,
for any column index $l$, there is at most one coefficient matrix $\Fbm_i$ with the $l^{th}$ column being non-zero.

\begin{figure*}[!tb]
	\begin{subfigure}[b]{0.22\textwidth}
		\includegraphics[]{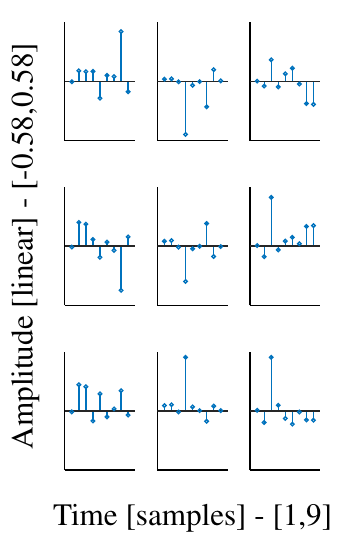}
		\caption{$\Dense_{\diter-1}(z)$}
		\label{fig:FFM_randomDense1}
	\end{subfigure}
	\hspace{0.2cm}
	\begin{subfigure}[b]{0.38\textwidth}
		\includegraphics[]{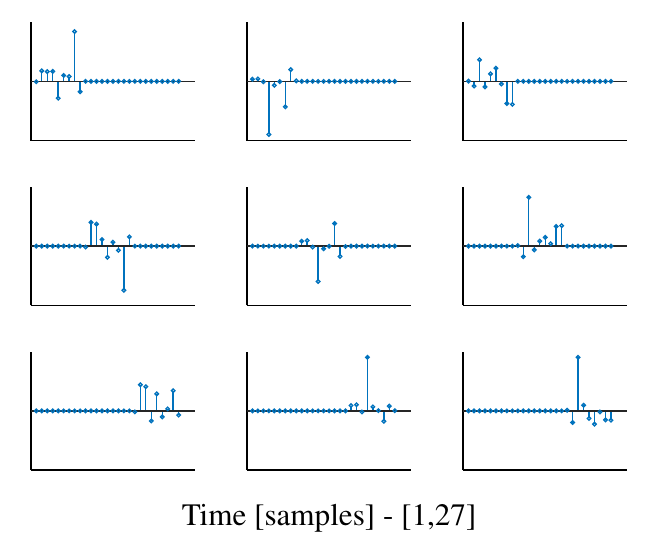}
		\caption{$ \stdDelayArgm{\Delay_{\diter}}{z} \Dense_{\diter-1}(z)$}
		\label{fig:FFM_randomDense2}
	\end{subfigure}
	\begin{subfigure}[b]{0.38\textwidth}
		\includegraphics[]{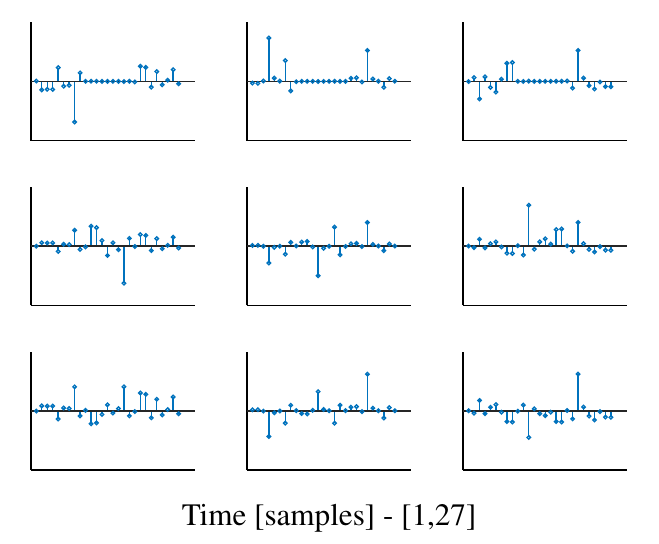}
		\caption{$\Dense_\diter(z) = \Unitary_{\diter} \stdDelayArgm{\Delay_{\diter}}{z}  \Dense_{\diter-1}(z)$}
		\label{fig:FFM_randomDense3}
	\end{subfigure}
	
	\caption{Random dense feedback matrices (RDFM) construction iteration from $\Dense_{\diter-1}(z)$ to $\Dense_{\diter}(z)$ for $\matSize = 3$. The FFM order of $\Dense_{\diter-1}(z)$ is $\filterOrder_{\diter-1} = 9$ and delay is $\Delay_{\diter} = [0,1,2] \cdot \filterOrder_{\diter-1} = [0, 9, 18]$.}
	\label{fig:FFM_randomDense_iteration}
\end{figure*}

We start the iteration with a dense unitary scalar matrix $\Dense_0(z) = \Unitary_0$, i.e., all matrix entries are non-zero. The corresponding filter order $\filterOrder_0  = 1$. 
With delays $\Delay_1 = [0,1,\dots,\matSize-1]$, $ \stdDelayArgm{\Delay_1}{z} \Dense_0(z)$ is column-distinct. For a dense unitary matrix $\Unitary_1$, the FFM
\begin{equation}
	\Dense_1(z) = \Unitary_1 \stdDelayArgm{\Delay_1}{z} \Dense_0(z)
	\label{eq:denseFirstIteration}
\end{equation}
is dense and the filter order is $\filterOrder_1  = \matSize$. The $\diter^{th}$ iteration is repeated analogously, however, with increased delays $\Delay_{\diter} = \filterOrder_{\diter-1} \Delay_1$. Thus, $\stdDelayArgm{\Delay_{\diter}}{z} \Dense_{\diter-1}(z)$ is column-distinct and for a dense unitary matrix $\Unitary_{\diter}$
\begin{equation}
	\Dense_\diter(z) = \Unitary_{\diter} \stdDelayArgm{\Delay_{\diter}}{z} \Dense_{\diter-1}(z)
	\label{eq:denseIteration}
\end{equation}
is dense and the filter order is $\filterOrder_\diter  = \matSize^{\diter}$. One iteration step of \eqref{eq:denseIteration} is depicted in Fig.~\ref{fig:FFM_randomDense_iteration}. Fig.~\ref{fig:FFM_randomDense} gives an example of a random dense feedback matrix (RDFM) with random unitary matrices $\Unitary_\diter$ and $\numStages = 3$ iterations. Each matrix element is therefore a dense FIR filter with order $\matSize^{\numStages} = 64$. Please note that the presented formulation is not unique and many permutations and variations may exist.

With a similar iterative procedure, we can construct a paraunitary Hadamard matrix $\HadaMat_\diter(z)$, i.e., all entries are of equal magnitude\footnote{The classic Hadamard matrix $\HadaMat$ has all entries either $+1$ or $-1$ and is orthogonal with $\HadaMat \HadaMat\tran = \matSize \eye$. For notational convenience, we use nonetheless the orthonormal definition and scale $\HadaMat$ such that $\HadaMat \HadaMat\tran = \eye$. Hadamard matrices do not exist for all $\matSize$, but for $\matSize = 1, 2, 4$ and $\matSize = 4k$ for most integers $k$.}. To this end, the unitary matrices in \eqref{eq:denseIteration} are replaced with scalar Hadamard matrices $\HadaMat$, i.e., orthogonal matrices with all entries equal to $\pm \matSize^{-1/2}$. Thus, $\HadaMat_0(z) = \HadaMat$ and 
$\stdDelayArgm{\Delay_{\diter}}{z} \HadaMat_{\diter-1}(z)$
is column-distinct and 
\begin{equation}
\HadaMat_\diter(z) = \HadaMat \stdDelayArgm{\Delay_{\diter}}{z}  \HadaMat_{\diter-1}(z) 
\label{eq:HadamardIteration}	
\end{equation}
is dense. Now, we show that $\HadaMat_\diter(z)$ is also Hadamard. Because, $\stdDelayArgm{\Delay_{\diter}}{z} \HadaMat_{\diter-1}(z)$ is column-distinct, any coefficient matrix has at most one non-zero column. In fact, due to the minimal delays, each such non-zero column has only entries with $\pm \matSize^{-\diter/2}$. Thus, the multiplication of such a coefficient matrix with the Hadamard matrix results in a dense matrix that has only entries $\pm \matSize^{-(\diter+1)/2}$.

\subsection{Sparse FIR FFM and Paraunitary Velvet} 
\label{sec:velvetFFM}
From perceptual studies of velvet noise sequences \cite{Valimaki:2013ft}, it is known that reverberation tails do not necessarily require to be fully dense. In general, a few pulses per millisecond are considered to be sufficient \cite{Alary:2017ug, Schlecht:2018ue}. Inspired by velvet noise sequences, we propose FIR FFM for a given density of $\velvetDensity$, i.e., the average number of pulses per sample (counted for each filter individually). In this sense, the dense FIR FFM corresponds to a density of $\velvetDensity = 1$.     

The construction in the preceding section can be readily adapted to incorporate density. Instead of using the smallest possible shift, we choose distinct values $\Delay_{1} \in [0,(\matSize-1)/\velvetDensity]$ and set $\filterOrder_\diter =  \lceil \matSize^{\diter} / \velvetDensity \rceil$. Instead of choosing fully random values for $\Delay_{1}$, some regularity can be enforced to generate evenly distributed pulses \cite{Jarvelainen:2007wp}. To preserve the initial density, the subsequent delays are unaltered, i.e., $\Delay_{\diter} \approx \filterOrder_{\diter-1} \Delay_1$ with little variations to promote irregularity. Similar to \eqref{eq:HadamardIteration}, the sparse FFM can be based on Hadamard matrices such that all matrix entries are either of the same magnitude or zero. For this reason, we like to call this special sparse FFM, a velvet feedback matrix (VFM). In Fig.~\ref{fig:FFM_velvet}, a VFM for $\matSize = 4$ and $\numStages = 2$ is depicted. Each matrix entry is a velvet sequence with $N^{\numStages} = 16$ pulses.

\subsection{Implementation and Computational Complexity}

In this section, we discuss the practical implementation of FFMs and the associated computational costs. For IIR FFM, the authors are not aware of a general way to accelerate implementation other than evaluating each IIR filter entry separately. Because of this reason, IIR FFM is considered rather computationally expensive for higher-order dense IIR FFMs. However, for FIR FFMs, we propose two approaches to implement the filtering efficiently: Firstly, by fast convolution and, secondly, cascaded multiplication. Further, any efficient unitary matrix multiplication like the Hadamard transform, Householder transform, and Circulant matrices are applicable similarly in the FFM context. 

\subsubsection{Fast Convolution}
As $\Fbm(z)$ consists only of FIR filters, the path from every input to every output can be implemented by fast convolution. However, it is possible to reduce the necessary $2\matSize^2$ FFT transforms to $2\matSize$ by processing the matrix mixing in the frequency domain. Every input signal is transformed into the frequency domain by an FFT, the complex vector is processed by a scalar complex matrix multiplication and then again back-transformed to the time domain by an inverse FFT. Compared to the scalar matrix $\Fbm$, the computational costs only increase by the forward and backward FFT, which depends on the lengths of the FIRs in $\Fbm(z)$. The computational complexity of the fast convolution for an FIR FFM with length $\filterOrder$ and size $\matSize \times \matSize$ is $\BigO( \matSize \filterOrder \log \filterOrder ) $ for the FFT and $\matSize^2$ for the complex matrix multiplication.


\begin{table*}[!tb]
\caption{Number of operations and filter order for paraunitary Hadamard matrix of size $\matSize \times \matSize$ with $\numStages$ stages. The number of operations is identical for the velvet feedback matrix (VFM). For comparison, a scalar matrix with $\matSize = 16$ is given. \schlecsn{Read and write operations to the main delays were omitted as they are included for the signal input and output operations.}}
\centering\ra{1.3}
\begin{tabular}{@{}rrrrcrrrcrrrcrr@{}}
\toprule& 
\multicolumn{3}{c}{$\matSize = 2$} & \phantom{abc}& \multicolumn{3}{c}{$\matSize = 4$} & \phantom{abc} & \multicolumn{3}{c}{$\matSize = 8$} & \phantom{abc} & \multicolumn{2}{c}{$\matSize = 16$}
\\\cmidrule{2-4} \cmidrule{6-8} \cmidrule{10-12} \cmidrule{14-15} & $\numStages=2$ & $\numStages=3$ & $\numStages=4$ && $\numStages=2$ & $\numStages=3$ & $\numStages=4$ && $\numStages=2$ & $\numStages=3$ & $\numStages=4$ && Scalar
\\\midrule
Addition \& Subtraction & 4 & 6 & 8 && 16 & 24 & 32 && 48 & 72 & 96 && 256 \\
Multiplication & 2 & 2 & 2 && 4 & 4 & 4 && 8 & 8 & 8  && 256 \\
Delay Read \& Write & 12 & 16 & 20 && 24 & 32 & 40 && 48 & 64 & 80 && 0 \\
Pulses per Filter & 4 & 16 & 64 && 16 & 256 & 4096 && 64 & 4096 & 262144 && 1  
\\\bottomrule
\end{tabular}
\label{tab:VFMoperations}
\end{table*}

\subsubsection{Cascade Multiplication}
Alternatively, the FIR FFM can directly implement the cascaded form in \eqref{eq:FFMfactorloose} with alternating processing of delays $\stdDelayArgm{\Delay_{\diter}}{z}$ and mixing $\Unitary_{\diter}$. The diagonal delay matrix $\stdDelayArgm{\Delay_{\diter}}{z}$, implemented with ring buffers, require $\matSize$ delay read and write operations plus circular pointer shifts. \schlecsn{The operations count for $\numStages$ stages in Fig.~\ref{fig:paraunitaryGeneralFIR}, is $2 \numStages \matSize$ for the delay matrices and $\numStages \matSize^2$ for the matrix multiplications. We want to mention the computational cost of the VFM specifically. The Hadamard transform, i.e., the multiplication with the fast Walsh-Hadamard transform, can be realized by $\matSize \log \matSize$ addition and subtractions.} In the VFM, the scaling factors of all Hadamard matrices can be summarized at the end of the VFM and realized by a total of $\matSize$ multiplication. See Table~\ref{tab:VFMoperations} for examples of operations count both for the paraunitary Hadamard matrices and VFMs. For comparison, Table~\ref{tab:VFMoperations} also lists the operations count of a scalar matrix multiplication with $\matSize = 16$. Besides that, the FFM of smaller matrix size $\matSize$ has fewer operations in general; importantly also other operations such as attenuation filters get reduced with the smaller matrix sizes.


%

\section{Modal Distribution}
\label{sec:modalDistribution}
In this section, we present the effect of FFMs on the modal distribution of the FFDN. First, we propose a method for lumped attenuation filters, which we subsequently evaluate by computing the modal decay distribution. In \cite{Schlecht:2019uj}, we have developed a large-scale modal decomposition technique for scalar feedback matrices $\Fbm$. In the following, we outline the general method and give an extension to FFMs $\Fbm(z)$.

\subsection{Attenuation Filters}
In Section~\ref{sec:lossyFeedback}, we introduced the lossy FFDN with the global gain-per-sample $\gain(z)$ related to reverberation time by \eqref{eq:attenuationRT60}. For a frequency-independent gain $\gain$, the criterion in \eqref{eq:lumpedAttenuationCriteria} can be satisfied strictly
\begin{equation}
	\stdDelayArg{ \frac{\gain}{z} } - \Unitary \p*{\frac{z}{\gain}} = \stdDelayArg{ \inv{z} } \diag{ \gain^{-\Delay} } - \Unitary_{\gain} \p*{z},
	\label{eq:broadbandAttenuationFFM}
\end{equation}
where
\begin{equation*}
	\Unitary_{\gain}(z) = \diag{ \gain^{\Delay_0}} \stdDelayArgm{\Delay_0}{z} \prod_{\diter=1}^\numStages \Unitary_\diter \diag{ \gain^{\Delay_\diter} } \stdDelayArgm{\Delay_\diter}{z}.
\end{equation*}
Thus, by pre-multiplying $\Unitary_\diter$ by $\diag{\gain^{\Delay_\diter}}$ in the FFM stages, no additional filtering operations need to be performed. However for frequency-dependent gain $\gain(z)$, this approach adds a large number of extra filters. Instead, we propose a solution to \eqref{eq:lumpedAttenuationCriteria} with a single lumped filter matrix $\GainMat(z)$. We extend the solution for the standard FDN \eqref{eq:attenuationFDN} to   
\begin{align}
\abs{\GainMat(\ejw)} = \diag{\gain(\omega)^{\Delay + \GroupDelayL(\omega) + \GroupDelayR(\omega) }},
	\label{eq:damping}
\end{align} 
where the left and right group delay, $\GroupDelayL(\omega)$ and $\GroupDelayR(\omega)$ respectively, approximate the FFM group delay, i.e., 
\begin{equation}
	\min_{\GroupDelayL,\GroupDelayR} \,   \anynorm*{ \GroupDelayL(\omega) \outerSum \GroupDelayR\tran(\omega) - \GroupDelayA(\omega)},
	\label{eq:groupDelay}
\end{equation}
where $\outerSum$ denotes the outer sum, i.e., $\p*{\vec{u} \outerSum \vec{v}\tran}_{ij} = u_i + v_j$ and 
\begin{equation}
	\GroupDelayA(\omega) = \dv{\arg \Fbm(\ejw)}{\omega}
\end{equation}
is the matrix of group delays. \schlecsn{For example, with the delay feedback matrix (DFM) \eqref{eq:DFMdefine}, we have 
\begin{align}
	\GroupDelayA(\omega) &= \dv{}{\omega} \arg \p*{\stdDelayArgm{\Delay_1}{\ejw} \Unitary_1 \stdDelayArgm{\Delay_0}{\ejw}} \\
	&=  \dv{}{\omega} \p*{ \Delay_1 \omega \outerSum \Delay_0\tran \omega + \arg \Unitary_1 } \\
	&= \Delay_1 \outerSum \Delay_0\tran
\end{align}
The minimization in \eqref{eq:groupDelay} yields $\GroupDelayL \equiv \Delay_1$ and $\GroupDelayR \equiv \Delay_0$ with no approximation error.
In Fig.~\ref{fig:FFM_delayMatrix}, we have $\Delay_1 \equiv [6, 0, 7, 5]$ and $\Delay_0 \equiv [ 12, 8, 0, 2]$ such that
\begin{equation}
	\GroupDelayA = \begin{bmatrix}
	  18  &  14  &   6   &  8 \\
    12  &   8  &   0   &  2 \\
    19  &  15  &   7   &  9 \\
    17  &  13  &   5   &  7 \\
    \end{bmatrix}
\end{equation}
which corresponds to the position of the non-zero sample for each FIR filter in the matrix. A general solution to obtain  $\GroupDelayL(\omega)$ and $\GroupDelayR(\omega)$ is outlined in Appendix~\ref{sec:groupDelayAppendix}.}

Any approximation error in \eqref{eq:groupDelay} causes the modal decay to deviate from the specified reverberation time. In the following, we quantify the decay deviation by performing a modal decomposition.

\subsection{Modal Decomposition}
The modal decomposition of FFDN computes the partial fraction decomposition of the transfer function in \eqref{eq:DFMdefine}, i.e., 
\begin{equation}
	H(z) = \directgain + \sum_{\poleIndex=1}^\N \frac{\residue_\poleIndex}{1 - \pole_\poleIndex \,z^{-1}},
	\label{eq:modalDecomposition}
\end{equation}
where $\residue_\poleIndex$ is the residue of the pole $\pole_\poleIndex$. The modal decomposition in \eqref{eq:modalDecomposition} can be computed even for high orders with the polynomial Ehrlich-Aberth Method. The method starts with a set of initial poles $\Pole\iter{0} \subset \set{C}$, which are typically placed along the unit circle. The Ehrlich-Aberth Iteration (EAI) provides the sequence of estimates
 \begin{equation}
	\pole_\poleIndex\iter{\iterIndex+1} = \pole_\poleIndex\iter{\iterIndex} - \EAIstep{\poleIndex}{\iterIndex}
	\label{eq:EAI}
\end{equation}
with the EAI step being
\begin{equation}
	\EAIstep{\poleIndex}{\iterIndex} = \frac{1}{{\inv{\newton{\pole_\poleIndex\iter{\iterIndex}}} -  \deflation{\poleIndex}{\Pole\iter{\iterIndex}}}}.
	\label{eq:EAIsimple}
\end{equation}
The Newton correction term is
\begin{equation}
	\newton{z} = \frac{\gcp(z)}{\gcp'(z)} = \frac{1}{ z^{-1}\N - z^{-2}\frac{\gcpR'\pzinv}{\gcpR\pzinv} }
	\label{eq:newtonCorrection}
\end{equation} 
and the deflation term is
\begin{equation}
	\deflation{\poleIndex}{\Pole\iter{\iterIndex}} = \sum_{l=1, l\neq \poleIndex}^\N \frac{1}{\pole_\poleIndex\iter{\iterIndex} - \pole_l\iter{\iterIndex}}
	\label{eq:Deflation}
\end{equation}
and $\gcpR(z) = z^\N \gcp \pzinv$ indicates the reversed GCP. The deflation term may be interpreted as a penalty term if two eigenvalues approach each other too closely and guarantee that all eigenvalues reached are unique. For numerically stable evaluation of Newton correction in \eqref{eq:newtonCorrection}, the left side is evaluated for $\abs{z} \leq 1$ and the right side otherwise \cite{Bini:1996fk}.    

For the FFM, we give the explicit form of both Newton correction terms. The corresponding derivations can be found in  Appendix~\ref{sec:modalDecompositionAppendix}. The inverse Newton correction for a FFM $\Fbm(z)$ is   
\begin{equation}
	\frac{\gcp'(z)}{\gcp(z)} = \trace \p*{ \MatPoly(z)^{-1} \, \MatPoly'(z) } + \frac{\deg \Fbm(z)}{z}.
	\label{eq:newtonCorrectionReversed}
\end{equation}    
The reversed GCP is
\begin{equation}  
	\gcpR(z) = (-1)^\systemOrder \detp{\Fbm\pzinv} \detp{ \MatPolyR(z) },
\end{equation}
where
\begin{equation}  
	\MatPolyR(z) =  \stdDelay^{-1} - \inv{\Fbm}\pfzinv.
\end{equation}
Similar to \eqref{eq:newtonCorrection}, we have
\begin{equation}  
	\frac{\gcpR'(z)}{\gcpR(z)} = \trace \p*{ \MatPolyR(z)^{-1} \, \MatPolyR'(z) - z^2 \inv{\Fbm}\pfzinv \, \Fbm'\pfzinv}.
	\label{eq:reversedNewtonCorreciton}
\end{equation}
The matrix inverse $\inv{\Fbm}\pzinv$ is challenging to compute in general. However, given the cascaded representation in \eqref{eq:FFMfactorloose} allows to do a matrix inversion of each cascaded component by itself, i.e.,
\begin{equation}
	\inv{\Fbm}\pzinv = \stdDelayArgm{\Delay_0}{z} \inv{\Unitary}_1  \cdots \inv{\Unitary}_\numStages \stdDelayArgm{\Delay_\numStages}{z}.
	\label{eq:FFMinverse}
\end{equation}
In case of a paraunitary FFM, computing the inverse is trivial as $\inv{\Fbm}\pzinv = \Fbm\tran(z)$.


\begin{figure}[!tb]
\includegraphics{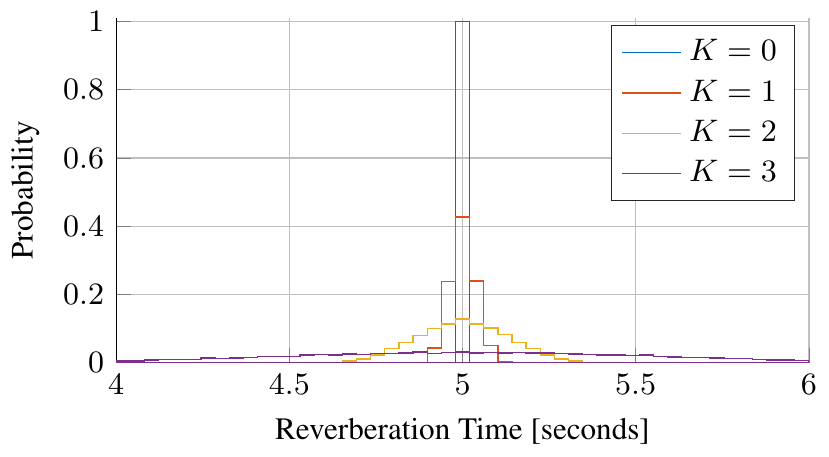}
	\caption{Modal decay distribution for VFMs with density $\velvetDensity = 1/30$ and $\matSize = 4$. Therefore, filter length is $\filterOrder \approx 30, 120, 480, 1920$ for $\numStages = 0,1,2,3$. The VFM is similar to Fig.~\ref{fig:FFM_velvet} with four main delays and a total system order of $\N \approx 10.000$ with reverberation time target of 5~seconds.}
	\label{fig:scatteringDecayDistribution}
\end{figure}

\subsection{Modal Decay Distribution}

In general, the larger the group delay in the FFM $\Fbm(z)$, the larger is the possible error in \eqref{eq:groupDelay} and therefore the deviation in modal decay. In the following, we give an example case to study the modal decay distribution. Consider a FFDN with $\matSize = 4$ and VFMs with density $\velvetDensity = 1/30$ (see Section~\ref{sec:velvetFFM}). Therefore, the filter order of the VFMs is $\filterOrder \approx 30, 120, 480, 1920$ for $\numStages = 0,1,2,3$. The main delays are $\Delay \in \bracket{ 1.000 , 5.000 } $ such that the total system order \eqref{eq:totalSystemOrder} is $\N \approx 10.000$. The target frequency-independent reverberation time is 5~seconds. Although, this can be solved accurately with \eqref{eq:broadbandAttenuationFFM}, we attempt to solve it with the lumped formulation \eqref{eq:damping} to assess the frequency-dependent case.

Fig.~\ref{fig:scatteringDecayDistribution} shows the probability distribution of modal decay for various sparse VFMs. Expectedly, the modal decay distribution becomes wider for longer filter length $\filterOrder$. The mean absolute group delay error between $\GroupDelayL(\omega) \outerSum \GroupDelayR\tran(\omega)$ and $\GroupDelayA(\omega)$ are about $0, 2, 10, 40$ samples for $\numStages = 0,1,2,3$, respectively. The maximum reverberation time error is below $0\%, 2\%, 8\%, 20\%$ for $\numStages = 0,1,2,3$, which is acceptable compared to the JND of $5\%$ for reverberation time \cite{Seraphim:1958um, Prawda:2019tq}. 


%
%

\begin{figure}[!tb]
	\includegraphics{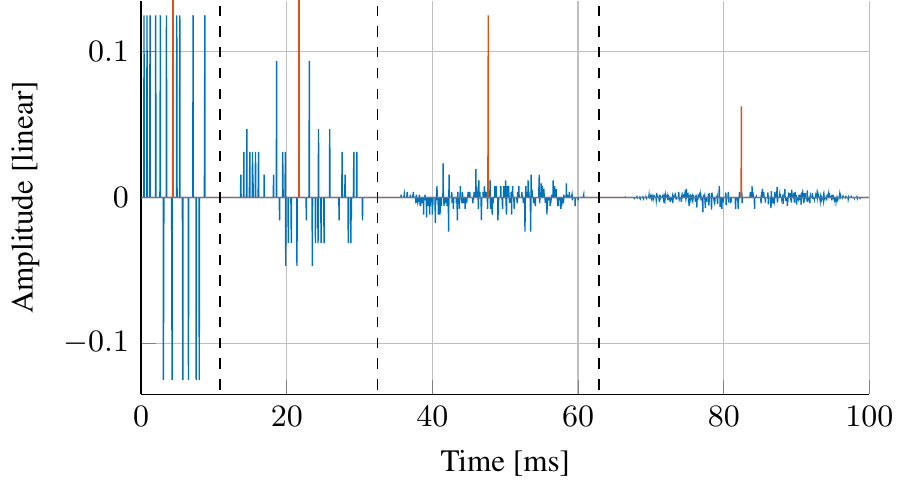}
	\caption{Evolution of a reflection path $\echoTF_{\Paths}(z)$ compared VFM (blue) and a standard FDN (red). The dashed line indicates the boundaries between each reflection. The first pulse amplitudes for the standard FDN are $0.5$ and $0.25$, but for better readability, the value range is limited to $\pm0.13$. \schlecsn{The cumulative temporal spreading relates to the scattering-like effect caused by incoherent reflections (see Fig.~\ref{fig:scattering}).}}
	\label{fig:scatterEvolution}
\end{figure}

\begin{figure}[!tb]
	\begin{subfigure}[b]{0.49\textwidth}
		\includegraphics[]{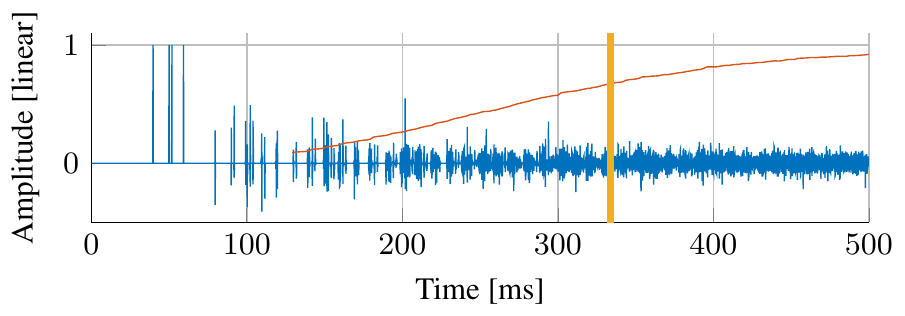}
		\caption{Elemental block feedback matrix (EBFM) with $\numStages = 64$ in \eqref{eq:FFMfactorization}.}
		\label{fig:IR_FFM_elemental64}
	\end{subfigure}
	\vspace{-0.1cm}
	\begin{subfigure}[b]{0.49\textwidth}
		\includegraphics[]{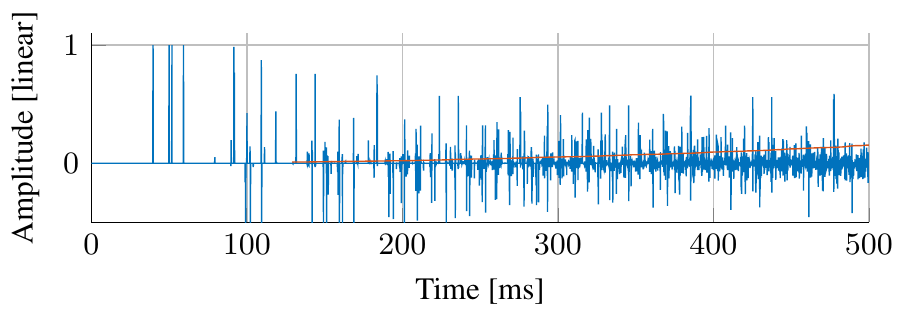}
		\caption{Delay feedback matrix (DFM) with $\numStages = 1$ in \eqref{eq:DFMdefine}.}
		\label{fig:IR_FFM_delayMatrix}
	\end{subfigure}
	\vspace{-0.1cm}
	\begin{subfigure}[b]{0.49\textwidth}
		\includegraphics[]{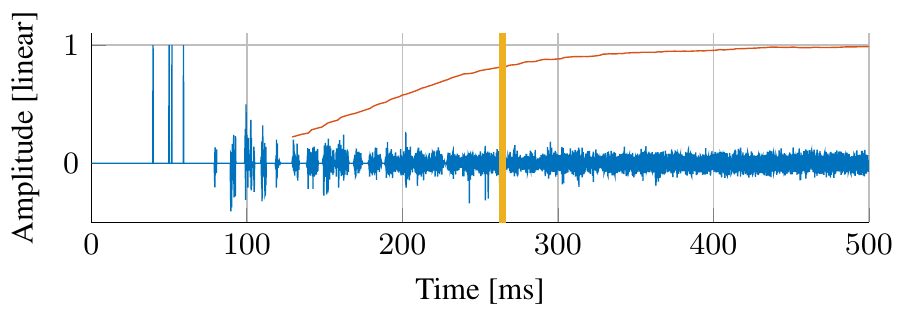}
		\caption{Random dense feedback matrix (RDFM) with $\numStages = 3$ in \eqref{eq:denseIteration}.}
		\label{fig:IR_FFM_randomDense}
	\end{subfigure}
	\vspace{-0.1cm}
	\begin{subfigure}[b]{0.49\textwidth}
		\includegraphics[]{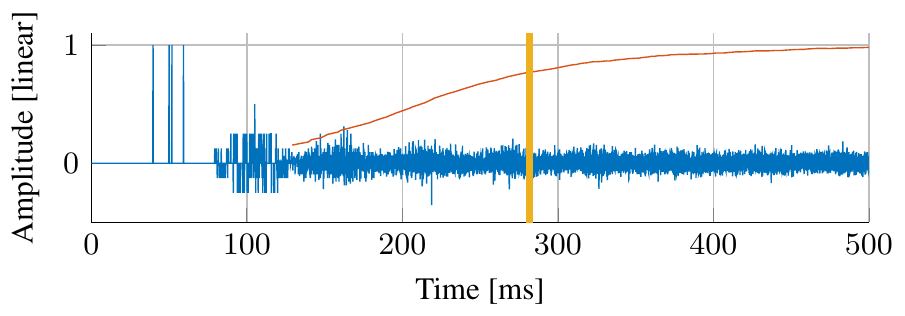}
		\caption{Velvet feedback matrix (VFM) with $\numStages = 2$ and $\velvetDensity = \frac{1}{30}$ in \eqref{eq:HadamardIteration}.}
		\label{fig:IR_FFM_velvet}
	\end{subfigure}
	\vspace{-0.1cm}
	\begin{subfigure}[b]{0.49\textwidth}
		\includegraphics[]{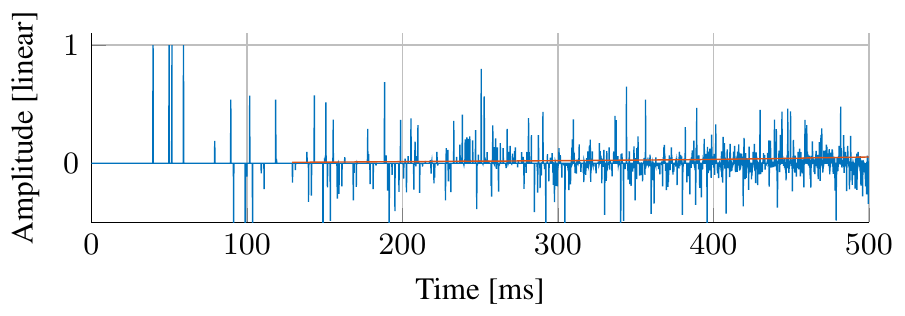}
		\caption{Scalar feedback matrix with $\matSize = 4$ (SFM-4).}
		\label{fig:IR_FFM_scalarMatrix}
	\end{subfigure}
	\vspace{-0.1cm}
	\begin{subfigure}[b]{0.49\textwidth}
		\includegraphics[]{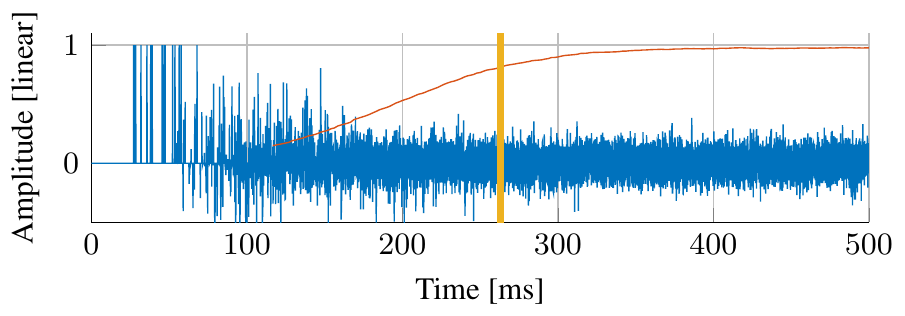}
		\caption{Scalar feedback matrix with $\matSize = 16$ (SFM-16).}
		\label{fig:IR_FFM_scalarMatrix16}
	\end{subfigure}
	\caption{The impulse response of FFDNs with FFMs in Fig.~\ref{fig:FFM}. The red line indicates the echo density profile, while the mixing time predicted using \cite{Tukuljac:2019ho} is shown in yellow.}
	\label{fig:IR_FFM}
\end{figure}

\section{Temporal Density}
\label{sec:temporalDensity}

In this section, we investigate the temporal density of the impulse response generated by the proposed FFDN.  

\subsection{Echoes and Echo Paths}
We first consider \emph{echo paths}, which are paths through the FFDN. More formally, the echo path $\Paths \in \bracket{1,\dots,\matSize}^\pathLength$ of length $\pathLength$ describes a sequence of delay lines, which is traversed by an impulse input. The resulting \emph{echo} is the system response if considering only the corresponding echo path $\Paths$
\begin{equation}
	\echoTF_{\Paths}(z) = \outgain_{\paths_\pathLength} z^{-\delay_{\paths_\pathLength}} \paren*{\prod_{\iterIndex = 1}^{\pathLength-1} z^{-\delay_{\paths_{\iterIndex}}} \fbm_{\paths_{\iterIndex+1},\paths_\iterIndex}(z)} \ingain_{\paths_1}.
\end{equation}    
In this sense, introducing FFMs into the FFDN does, in general, not change the number and overall temporal distribution of the echoes. However, most importantly, it changes the temporal spread of the echoes themselves. Fig.~\ref{fig:scatterEvolution} shows the echoes through a VFM (depicted in Fig.~\ref{fig:FFM_velvet}) with $\pathLength = 2, 3, 4$ and $5$. While the first pass-through the VFM results in a sparse echo response, the more often an echo traverses the VFM, it also becomes denser. The temporal extent of the spread is roughly the filter order of the VFM multiplied by the number of matrix pass-throughs. The cumulative spreading is also in accord with physical rooms where each reflection at a boundary introduces scattering (see Fig.~\ref{fig:scattering}). This is in stark contrast to the standard FDN with a scalar feedback matrix, where any echo response is an impulse, and no spreading in time occurs. This illustrates a key feature of the proposed method: while in standard FDN high echo density requires more delay lines \cite{Schlecht:2017il}, the FFDN can balance the number of echoes and their temporal extent. In particular, the echo density can be controlled by the number of mix and delay stages and the delays in the FFM.


\begin{figure}[!tb]
	\includegraphics{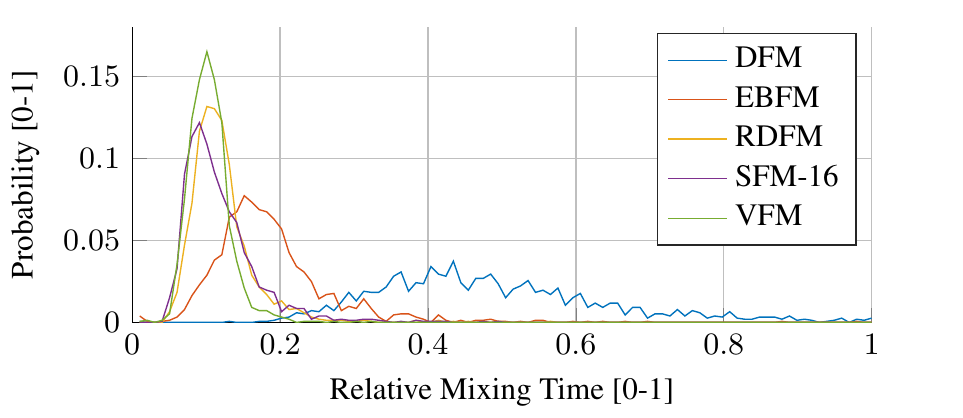}
	\caption{Probability distribution of the relative mixing time for FFM of type $X$, i.e., $\textrm{mixingTime}(X) / \textrm{mixingTime}(\textrm{SFM-4})$.}
	\label{fig:plot_scatteringDensity}
\end{figure}

\subsection{Echo Density Measure and Mixing Time}

We employ a recently proposed echo density measure based on a smooth sorted density measure \cite{Tukuljac:2019ho}. The echo density of the impulse response $\tf{}(n)$ is computed using a two-pass procedure. The input impulse response is converted to an echogram, i.e., $e(n) \equiv \tf{}^2(n)$. A local energy normalization factors out the energy decay envelope. The normalized echogram is analyzed with a rectangular sliding window centered at each sample. Then the sorted density is computed as a fraction of the window width. Processing for each sample and normalizing with the expected value for Gaussian noise yields echo density. This is fitted with a general power-law model motivated by theoretical considerations. Overall, it has been shown to provide a smoother and more reliable echo density curve compared to the model of Abel and Huang proposed in \cite{Schlecht:2017il, Abel:2006vl, Huang:2007ul, Lindau:2012vi}.


In Fig.~\ref{fig:IR_FFM}, the impulse responses of six FFDNs are plotted alongside their echo density profile. Four of the FFDNs are the same as in Fig.~\ref{fig:FFM}, and two additional standard FDNs with $\matSize = 4$ and $\matSize = 16$ for comparison. The four matrix types EBFM, RDFM, VFM, SFM-16 were parametrized such that the impulse responses mix at similar times, DFM and SFM do not mix within the first 0.5~seconds. In the following, we quantify the mixing time of FFDN via a Monte Carlo simulation.

We generate random variations of the six FFDN types as presented in Fig.~\ref{fig:IR_FFM}: EBFM, DFM, RDFM, VFM, SFM-4, and SFM-16. The main delays are generated between $1000$ and $8000$ samples, and the FFM were generated with random seeds as described in Section~\ref{sec:FFM}. We use the SFM-4 as the baseline, as it has the highest mixing time typically. Thus, the relative mixing time is the ratio between the evaluated mixing time and the baseline mixing time of SFM-4. Fig.~\ref{fig:plot_scatteringDensity} shows the probability distribution of the relative mixing time. The DFM reduces the mixing time by a factor of 2, while the EBFM reduces it by a factor of 5. The VFM and RDFM reduce the mixing time by a factor of 10 compared to the SFM-16. \schlecsn{Further, the computational cost of the VFM is only a fraction of the SFM-16 cost} (see Table~\ref{tab:VFMoperations}).

\section{Conclusion}
We proposed a novel artificial reverberation method, the filtered feedback delay network (FFDN), by introducing a scattering-like recursive filter to improve the echo density. Among the various designs for the filter feedback matrix (FFM), the velvet feedback matrix (VFM) is the most prominent innovation for its superior computational cost and induced echo density. A generalized design for attenuation filters in FFDNs was proposed. The effect of FFMs on the modal distribution was investigated by an extended modal decomposition method. The attenuation filter design keeps the modal decay deviation close to the JND of 5\%. Similarly, the echo density was investigated for various FFM designs. We showed that highly efficient FFM designs can reduce the mixing time by a factor of 10.  Thus, the FFDN solves the trade-off of standard FDN design between computational complexity, mode density, and echo density by introducing a computationally efficient scattering technique. The FFDNs render for the first time a low delay count viable for high-quality artificial reverberation. A possible future work is to connect the FFM design to a specified scattering effect, either derived from a perceptual or physical model.

\appendices

\section{Group Delay Approximation }
\label{sec:groupDelayAppendix}

One method to solve the minimization problem in \eqref{eq:groupDelay} is to convert it to a rank-1 approximation problem by applying an exponential function on both sides, i.e.,  
\begin{equation}
	\anynorm*{ \GroupDelayLexp(\omega) \GroupDelayRexp\tran(\omega) - \GroupDelayAexp(\omega) }, 
	\label{eq:rank1Approx}
\end{equation}
where $\GroupDelayLexp(\omega) = \exp( \GroupDelayL(\omega) )$, $\GroupDelayRexp(\omega) = \exp( \GroupDelayR(\omega) )$ and $\GroupDelayAexp(\omega) = \exp( \GroupDelayA(\omega) )$ and the exponential function is applied element-wise. The rank-1 approximation in \eqref{eq:rank1Approx} is then solved by taking the dominant vectors from a singular value decomposition of $\GroupDelayAexp(\omega)$. Applying a logarithm to $\GroupDelayLexp(\omega)$ and $\GroupDelayRexp(\omega)$ yields then a solution to the original problem. This is merely a practical solution that can be further improved depending on the desired error norm in \eqref{eq:groupDelay}.

\section{Modal Decomposition for FFDN }
\label{sec:modalDecompositionAppendix}

Given a FFM $\Fbm(z)$ and delays $\stdDelay$, the general characteristic polynomial is
 \begin{equation}
 	\gcp(z) = z^{\deg \Fbm(z)} \detp{\P}
 \end{equation}
and the corresponding derivative is
 \begin{equation*}
 	\gcp'(z) = z^{\deg \Fbm(z)} \detp{\P} \p*{ \trace \p*{ \MatPoly(z)^{-1} \, \MatPoly'(z) } + \frac{\deg \Fbm(z)}{z} }.
 \end{equation*}
The quotient of the previous two terms yields than \eqref{eq:newtonCorrectionReversed}. 

From \eqref{eq:totalSystemOrder}, we have $\sumDelay \coloneqq \sum_{\iterIndex = 1}^\matSize \delay_\iterIndex = \N - \deg \Fbm(z)$. The reversed GCP is (we write for convenience $\stdDelayn = \stdDelay$)
\begin{align*}
	\gcpR(z) &= z^\N \gcp \pzinv = z^{\N-\deg \Fbm(z)} \detp{\stdDelayn - \Fbm\pzinv} \\
	&= z^{\sumDelay} \detp{\stdDelayn}\detp{\Fbm\pzinv} \\ & \qquad \detp{\inv{\Fbm}\pzinv - \inv{\stdDelayn}} 
\end{align*}
and as $z^{\sumDelay} \detp{\stdDelayn} = \eye$, it is
\begin{align*}
	\gcpR(z) &= \detp{\Fbm\pzinv} \detp{-\eye}\detp{\inv{\stdDelayn} - \inv{\Fbm}\pzinv}\\
	&= (-1)^\matSize \detp{\Fbm\pzinv} \detp{\MatPolyR(z)}	 
\end{align*}
where
\begin{equation}  
	\MatPolyR(z) =  \stdDelayn^{-1} - \inv{\Fbm}\pzinv.
\end{equation}
Finally, the derivative of the reversed GCP is 
\begin{align*}
	\gcpR'(z)
	&=  (-1)^\matSize \detp{\Fbm\pzinv} \detp{\MatPolyR(z)}  \\
	& \qquad \qquad \trace \p*{ \MatPolyR(z)^{-1} \, \MatPolyR'(z) - z^2 \inv{\Fbm}\pzinv \, \Fbm'\pzinv}	 
\end{align*}
and the quotient of the previous equations yields the reversed Newton correction term in \eqref{eq:reversedNewtonCorreciton}.

\section*{Acknowledgment}

The authors would like to thank Nikunj Raghuvanshi and his colleagues for providing the MATLAB code of \cite{Tukuljac:2019ho}.

\bibliographystyle{IEEEtran}


\begin{IEEEbiography}[{\includegraphics[width=1in,height=1.25in,clip,keepaspectratio]{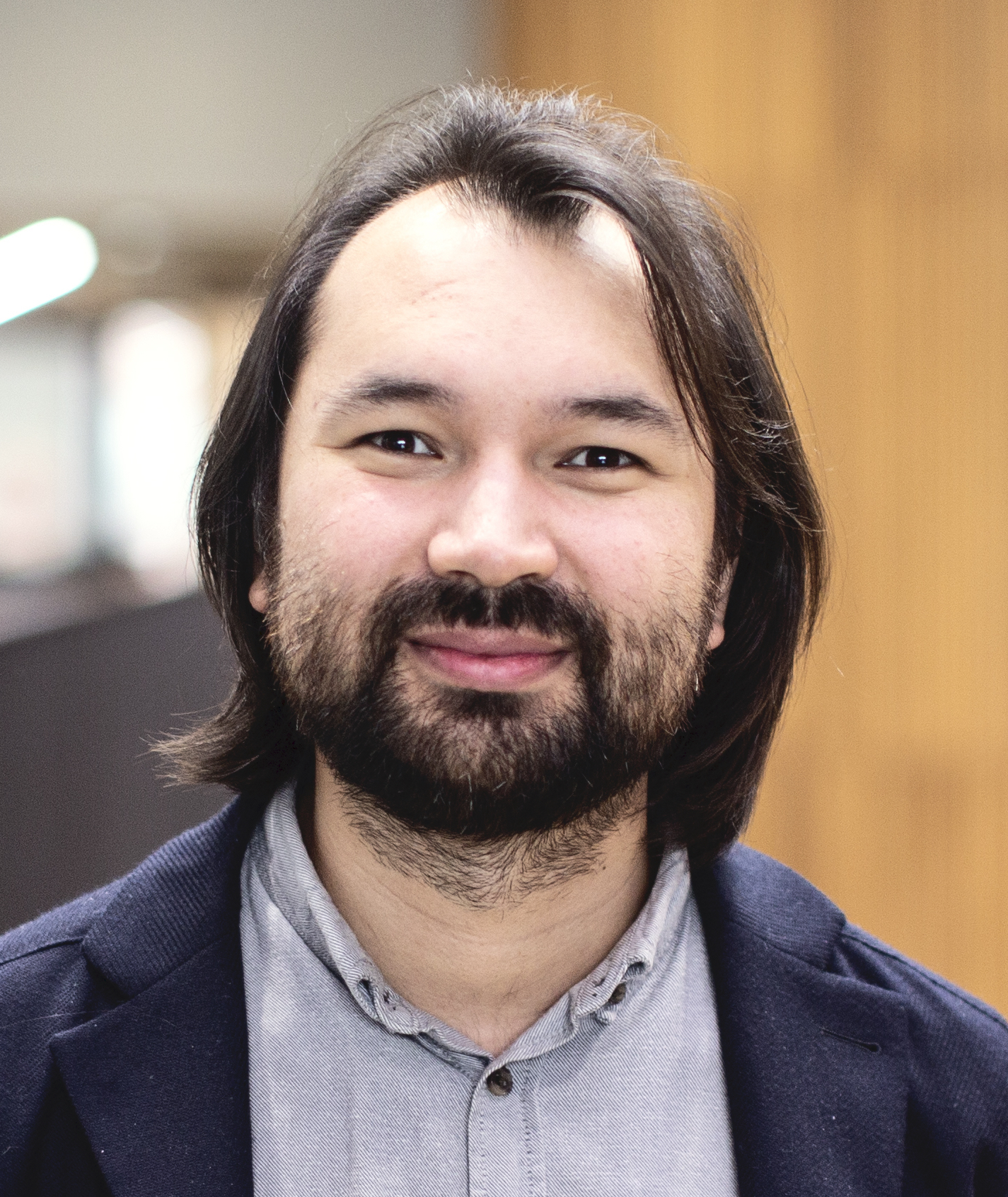}}]{Sebastian J. Schlecht} (M'19-SM'20) Sebastian J. Schlecht is a Professor of Practice for Sound in Virtual Reality at the Acoustics Labs, Department of Signal Processing and Acoustics and Media Labs, Department of Media, of Aalto University, Finland. He received the Diploma in Applied Mathematics from the University of Trier, Germany in 2010, and an M.Sc. degree in Digital Music Processing from School of Electronic Engineering and Computer Science at Queen Mary University of London, U.K. in 2011. In 2017, he received a Doctoral degree at the International Audio Laboratories Erlangen, Germany, on the topic of artificial spatial reverberation and reverberation enhancement systems. From 2012 on, Dr. Schlecht was also external research and development consultant and lead developer of the 3D Reverb algorithm at the Fraunhofer IIS, Erlangen, Germany. Since 2019, he is Professor of Practice for Sound in Virtual Reality at the Aalto University, Espoo, Finland, and affiliated with the Acoustics Lab and the Media Lab.   

His research interests are acoustic modeling and auditory perception of acoustics, analysis, and synthesis of feedback systems, music information retrieval, and virtual and augmented reality.

He is the recipient of multiple best paper awards including Best Paper in Journal of the Audio Engineering Society (JAES) in Jun 2020, Best Paper Award at IEEE Workshop on Applications of Signal Processing to Audio and Acoustics (WASPAA) in Oct 2019, 2nd Best Paper Award at International Conference on Digital Audio Effects (DAFx) in Sep 2018, and, Best Peer-Reviewed Paper at 2018 AES International Conference on Audio for Virtual and Augmented Reality (AES AVAR) in Aug 2018.\end{IEEEbiography}

\begin{IEEEbiography}[{\includegraphics[width=1.25in,height
=1.25in,clip,keepaspectratio]{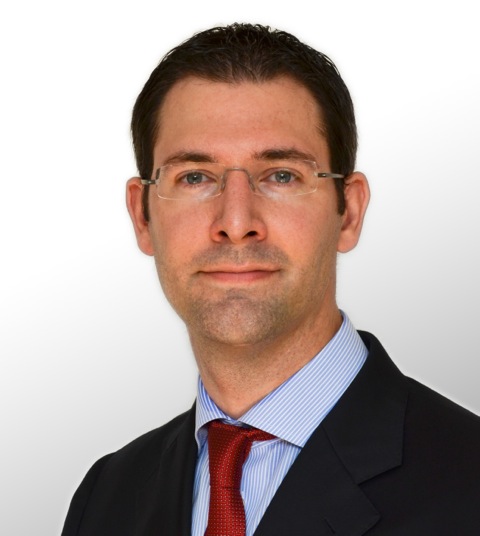}}]{Emanu\"{e}l A.P. Habets} (S'02-M'07-SM'11)
is an Associate Professor at the International Audio Laboratories Erlangen (a joint institution of the Friedrich-Alexander-Universit\"{a}t Erlangen-N\"{u}rnberg and Fraunhofer IIS), and Head of the Spatial Audio Research Group at Fraunhofer IIS, Germany. He received the B.Sc. degree in electrical engineering from the Hogeschool Limburg, The Netherlands, in 1999, and the M.Sc. and Ph.D. degrees in electrical engineering from the Technische Universiteit Eindhoven, The Netherlands, in 2002 and 2007, respectively. 

From 2007 until 2009, he was a Postdoctoral Fellow at the Technion - Israel Institute of Technology and at the Bar-Ilan University, Israel. From 2009 until 2010, he was a Research Fellow in the Communication and Signal Processing Group at Imperial College London, U.K. 

His research activities center around audio and acoustic signal processing, and include spatial audio signal processing, spatial sound recording and reproduction, speech enhancement (dereverberation, noise reduction, echo reduction), and sound localization and tracking.

Dr. Habets was a member of the organization committee of the 2005 International Workshop on Acoustic Echo and Noise Control (IWAENC) in Eindhoven, The Netherlands, a general co-chair of the 2013 International Workshop on Applications of Signal Processing to Audio and Acoustics (WASPAA) in New Paltz, New York, and general co-chair of the 2014 International Conference on Spatial Audio (ICSA) in Erlangen, Germany. He was a member of the IEEE Signal Processing Society Standing Committee on Industry Digital Signal Processing Technology (2013-2015), a Guest Editor for the IEEE Journal of Selected Topics in Signal Processing and the EURASIP Journal on Advances in Signal Processing, an Associate Editor of the IEEE Signal Processing Letters (2013-2017), and Editor in Chief of the EURASIP Journal on Audio, Speech, and Music Processing (2016-2018). He is the recipient, with S. Gannot and I. Cohen, of the 2014 IEEE Signal Processing Letters Best Paper Award. Currently, he is a member of the IEEE Signal Processing Society Technical Committee on Audio and Acoustic Signal Processing, a member of the EURASIP Technical Activities Board, and Chair of the EURASIP Technical Area Committee on Acoustic, Speech and Music Signal Processing.
\end{IEEEbiography}

\end{document}